\documentclass[prd,twocolumn,showpacs,reprint,preprintnumbers,nofootinbib]{revtex4-1}

\RequirePackage[colorlinks=true
,urlcolor=blue
,anchorcolor=blue
,citecolor=blue
,filecolor=blue
,linkcolor=blue
,menucolor=blue
,pagecolor=blue
,linktocpage=true
,pdfproducer=medialab
,pdfa=true
]{hyperref}

\usepackage{amsmath,amssymb,amsthm,amsfonts}
\usepackage{graphicx,tabularx}
\usepackage{color}
\usepackage{multirow}  
\usepackage{comment}

\makeatletter
\def\l@subsubsection#1#2{}
\makeatother

\newcommand{\be}{\begin{equation} \begin{aligned}}
\newcommand{\ee}{\end{aligned} \end{equation}}
\newcommand{\beqa}{\begin{eqnarray}}
\newcommand{\eeqa}{\end{eqnarray}}

\begin{document}
\title{On Profiles of Boson Stars with Self-Interactions}

\preprint{UCI-HEP-TR-2017-17}

\author{Felix Kling}
\email{fkling@uci.edu}
\affiliation{Department of Physics and Astronomy, University of
California, Irvine, CA 92697, USA}

\author{Arvind Rajaraman}
\email{arajaram@uci.edu}
\affiliation{Department of Physics and Astronomy, University of
California, Irvine, CA 92697, USA}

\begin{abstract}
{Under the influence of gravity, light scalar fields can form bound compact objects called boson stars. We use the semi-analytic approach of matching asymptotic expansions to obtain the profile for boson stars where the constituent particles have self-interactions. We obtain parametric representations of these profiles as a function of the self-interactions, including the case of very strong self-interactions. We show that our methods agree with solutions obtained by purely numerical methods. Significant distortions are found as compared to the noninteracting case.}
\end{abstract}

\maketitle
\setcounter{secnumdepth}{3}
\setcounter{tocdepth}{3}
\tableofcontents

\section{Introduction}

The increasingly strong constraints on weakly interacting massive particles have made axions more attractive as a dark matter candidate (for reviews see \cite{Kim:2008hd, Cheng:1987gp}). 
Many searches are ongoing to find axion-like particles \cite{1983PhRvL..51.1415S, 1985PhRvD..32.2988S, 
2010PhRvL.104d1301A, 2011PhRvD..84l1302H, Rybka:2014cya, Sikivie:2013laa, 2012PhRvD..85c5018B, Horns:2012jf, Budker:2013hfa, Graham:2013gfa, 1999NIMPA.425..480Z,Stadnik:2013raa, Feng:2018aa}. 

It has been noted by many authors that the axions might bind into compact spatial structures (e.g. see the review \cite{Liebling:2012fv}). These are more generally referred to as boson stars \cite{1968PhRv..172.1331K, Ruffini:1969qy,Carignano:2016lxe}). Such objects, if they exist, would produce distinctive signatures in axion search experiments, and understanding these signatures requires a description of the profile of the boson stars, which could produce unique time and spatial dependencies of the signal which can distinguish it from backgrounds \cite{JacksonKimball:2017qgk}. It is also important to know how these profiles are distorted by the presence of self-interactions, and also by the gravitational effects of matter. For all these reasons, it is timely to have a precise description of the profiles of boson stars.
 
In a previous communication~\cite{Kling:2017mif}, we analyzed the profile of such objects in the special case when the bosons had no self-interactions. These objects are solutions to a coupled set of equations called the Gross-Pitaevskii-Poisson equations. In that paper, we showed that one could find approximate  solutions to these equations, by using a combination of analytical and numerical methods. We showed that our methods were numerically stable, and that they converged uniformly far away from the core of the star, and furthermore were much less computationally expensive compared to other purely numerical methods \cite{Ruffini:1969qy, Barranco:2010ib, 2007JCAP...06..025B, 1989PhRvA..39.4207M, 1996PhRvD..53.2236L, 2000NewA....5..103G, 2003PhRvD..68b3511A,Chavanis:2011zm, Eby:2014fya,Eby:2015hsq,Yang:2015paa,Kan:2017uhj,Eby:2017teq,Visinelli:2017ooc,Chavanis:2017loo}. 

In this paper, we extend our previous methods to the case of interacting bosons. This is a more difficult situation, because the self-interactions can be much stronger than the gravitational binding. Nevertheless we show that our methods continue to provide excellent agreement with a fully numerical solution. 

Our results show that self-interactions can produce significant distortions of the profile of the star, as compared to the non-interacting star. We parametrize these deformations by finding a parametrization
for the star in terms of two matching asymptotic expansions, where the expansions are taken far away from the core and close to the core respectively. The parameters of the solution can be found systematically by matching the expansions in an overlap region. We find these parameters as a function of the interaction strength, which also allows us to find any desired physical quantity (mass, central density etc.) in terms of the coupling. We do this both for weak couplings, where we can perturb around the noninteracting star, as well as for strong couplings, where we can perturb around the Thomas-Fermi limit of the solution. We note that these expansions provide a solution to the axion stars which have an accuracy of at least $10^{-3}$, and can hence be used in lieu of complicated numerical calculations. 

Our results improve on other semi-analytic approaches (e.g. \cite{Chavanis:2011zi,Eby:2017teq, Schiappacasse:2017ham}), which have used variational and other techniques to find approximations to the boson star profile. Our methods are particularly suited to accurate evaluations of the profile away from the center, where the falloff is well described by a Whittaker function. 

In the following sections we rederive the Gross-Pitaevskii-Poisson equations satisfied by the boson star. We also describe various limiting cases, including the Thomas-Fermi limit of strong coupling. We then present a series expansion in the two asymptotic regimes, and find the solutions separately for weakly coupled systems and for strongly coupled systems. We show how our solutions apply to the special case of axion stars. We end with conclusions and directions for future work.

\section{Gross-Pitaevskii-Poisson Equations}

\subsection{The real scalar field}

In \cite{Kling:2017mif} we have derived the structure equations for the ground state of a self-gravitating complex scalar field in the non-relativistic limit. Following the procedure described in \cite{Braaten:2015eeu}, we now show that the same set of equations also apply for a real scalar field.

Let us consider the real scalar field $\phi(\vec{r},t)$ described by the Lagrangian
\be
\mathcal{L} =\frac{1}{2} g^{\mu\nu} (\partial_\mu \phi)(\partial_\nu \phi) - \frac{1}{2}m^2 \phi^2 - 
\frac{1}{12} \lambda \phi^4.
\label{gpp-Lagrangian}
\ee
In the presence of gravity, the scalar field can form gravitational bound states, which are called boson stars. A simple solution can be obtained assuming that only the ground state is populated. In this case the field can be expressed in terms of a single real function $\psi(r)$, sometimes called the wave function of the boson star, describing the radial profile of the boson star. We can write
\be
\phi (t,r) = \sqrt{\frac{N}{2E}} \psi(r) \left( e^{-i E t }+    e^{i E t} \right)
\label{gpp-field}
\ee
where $E$ is the ground state energy and $N$ is number of bosons in the ground state. Note that properly shifting the time coordinate allows us to absorb a possible phase of the wave function and therefore to choose $\psi$ to be real. We have chosen a wave function normalization $\int \psi^2 dV =1$ which allows us to identify $\psi^2$ with as the probability density. 

The equation of motion of the scalar field is the Klein-Gordon equation $\Box\phi+ m^2 \phi+ \frac{\lambda}{3} \phi^3=0$. Assuming that the field couples only weakly to gravity, we can use a Newtonian approximation. This allows us to introduce the Newtonian potential $\Phi$ in the metric $g_{\mu\nu}= \text{diag}(1+2\Phi,-1,-1,-1)$. We can then rewrite the Klein-Gordon equation as
\be
\frac{\partial^2_t \phi}{1+2\Phi} - \nabla^2 \phi + m^2 \phi+ \frac{\lambda}{3} \phi^3=0.
\ee 

Let us further assume that the ground state is non-relativistic. In this case we can write $E = m + e$ with binding energy $e \ll m$. This implies $e\psi, \Phi\psi, \nabla \psi \ll  m \psi $. Using $\partial_t^2 \phi = -E^2 \phi$ and $E=m+e$, we can rewrite the the Klein-Gordon equation in the non-relativistic limit as 
\be
- e \phi  -\frac{1}{2m} \nabla^2 \phi + m \Phi \phi + \frac{\lambda}{6 m} \phi^3  = 0 .
\ee

Inserting the explicit form of the field given in Eq.~\ref{gpp-field} and rephasing by $e^{iEt}$, we obtain the Schr{\"o}dinger-type equation 
\be
 e \psi =  -\frac{1}{2m} \nabla^2 \psi + m \Phi \psi + \frac{N \lambda }{4 m^2} \psi^3  .
\label{gpp-seq}
\ee
Here we have dropped additional terms with rapidly oscillating phase factor $e^{-inEt}$ where $n$ is a non-zero integer. For the non-relativistic approximation to be consistent, the last term should be sufficiently small, i.e. $\frac{N \lambda}{4 m^2} \ll m$.

The Newtonian potential is related to the energy density via the Poisson equation $\nabla^2 \Phi = 4\pi G\rho$. The energy density $\rho$ of the real scalar field is
\be
\rho =  \frac{1}{2} (\partial_t \phi)^2 + \frac{1}{2}  (\nabla \phi)^2+  \frac{1}{2} 
m^2 \phi^2 + \frac{\lambda }{12}\phi^4   \approx N m \psi^2
\ee
where we used the non-relativistic approximation in the last step. Newton's equation therefore takes the simple form 
\be
\nabla^2 \Phi = 4\pi G N m \psi^2
\label{gpp-neq}
\ee

Comparing with the results of \cite{Kling:2017mif}, we find that both the ground state of a boson star for both a complex scalar field and a real scalar field are described by the same set of equations given in Eq.~\ref{gpp-seq} and Eq.~\ref{gpp-neq}. These are often referred to as Gross-Pitaevskii-Poisson equations. Note that the Gross-Pitaevskii-Poisson equations for a real scalar field have also been obtained by the authors of \cite{Eby:2014fya}, which follow a semi-classical approach considering a quantized scalar field $\phi $. 

\subsection{Limits and Validity}
\label{gpp-validity}

The terms on the right-hand side of Eq.~\ref{gpp-seq} represent the contribution to the energy of a scalar particle due to the quantum pressure, gravity and classical pressure respectively. The quantum pressure is a consequence of Heisenberg's uncertainty principle and is always repulsive, preventing the star from gravitational collapse. Gravity on the other hand is always attractive. The classical pressure arises from he contact interaction term and can either be attractive or repulsive, depending on the sign of the self coupling $\lambda$. It is illustrative to consider the limits in which one of the three contributions is negligible. 

Using scaling relations between the coupling $\lambda$, the star's mass $M$ and the star's radius $R$, we will qualitatively discuss both physical properties of the star and the validity of the solution. Without providing a formal definition of the star's radius $R$, we know that the probability density inside the star scales like $\psi^2 \sim \frac{1}{R^3}$. Similarly, a field derivative will scale as $\nabla \psi = \frac{\psi}{R}$. For a more rigorous discussion of the mass-radius relations of boson stars see \cite{Chavanis:2011zi}.

\subsubsection*{Non-Interacting Limit $\lambda = 0$}

In the non-interacting case, $\lambda = 0$, the quantum pressure balances gravity. We have obtained a semi-analytical solution for this case in \cite{Kling:2017mif}. Note that for non-negligible couplings $\lambda \neq 0$, the boson star becomes effectively non-interacting at large radius due to low densities.

We can rewrite Eq.~\ref{gpp-seq} and \ref{gpp-neq} as
\be
4\pi GM\psi^2 = \nabla^2 \Phi = \frac{1}{2m^2} \nabla^2 \left(\frac{\nabla^2 \psi}{\psi}\right).
\ee
Using the scaling behavior discussed above, we see that the radius of a non-self-interacting boson star scales as $R \sim (GMm^2)^{-1} $. This is an remarkable result: the star's radius decreases when its mass increases. The binding energy of a scalar particle is 
\be
e \sim m\Phi \sim \frac{GMm}{R} \sim G^2 M^2 m^3 .
\ee
The non-relativistic approximation requires $e \ll m$ which implies $M \ll M_{\lambda=0}^{max} = \frac{M_{pl}^2}{m}$. 

\subsubsection* {Thomas-Fermi Limit}

For strong repulsive self-interactions $\lambda >\lambda_c$, the quantum pressure becomes negligible and the classical pressure balances gravity. Here $\lambda_c$ is the critical coupling at which the quantum pressure and classical pressure are equally important:
\be
\frac{1}{2m} \nabla^2 \psi = \frac{M\lambda_c}{4m^3} \psi^3.
\ee 
We can use the scaling behavior introduced earlier to solve for the coupling and obtain $\lambda_c \sim \frac{Rm^2}{M}$. At the critical coupling, hydrostatic equilibrium requires $R\sim(GMm^2)^{-1}$ which implies $\lambda_c \sim \frac{M_{pl}^2}{M^2}$. Already a very small coupling is sufficient for the Thomas-Fermi limit to apply.

For $\lambda \gg \lambda_c$ we can rewrite Eq.~\ref{gpp-seq} and \ref{gpp-neq} as
\be
4\pi GM\psi^2 = \nabla^2 \Phi = \frac{M}{4m^4} \nabla^2  \psi^2.
\ee
Using that the gradient scales as $\nabla\psi \sim \frac{\psi}{R}$, we obtain that the radius of a strongly self-interacting boson star scales as $R \sim \frac{M_{pl}}{m^2} \sqrt{\lambda}$. The radius is independent of the star's mass. This is not surprising, since both gravity and the repulsive self-interaction are proportional to the number of particles and therefore the star's mass. The binding energy of a scalar particle is 
\be
e \sim m\Phi \sim \frac{GMm}{R} \sim  \frac{M}{\sqrt{\lambda}}\frac{m^3}{M_{pl}^3}
\ee
The non-relativistic approximation requires $e \ll m$ which implies $M \ll M_{\lambda>0}^{max} = \sqrt{\lambda} \frac{M_{pl}^3}{m^2} = \sqrt{\lambda} \frac{M_{pl}}{m} M_{\lambda=0}^{max}$. Larger couplings increase the validity range of the non-relativistic approximation to higher masses $M$.

An analytic solution for the Thomas-Fermi limit has been obtained in \cite{2007JCAP...06..025B} and is discussed in Sec.~\ref{sec-tfl}.

\subsubsection*{Non-Gravitational Limit }

For sufficiently strong attractive self-interactions, $\lambda < \lambda^*$, quantum pressure balances the attractive self-interaction while the effect of gravity becomes negligible. Gravity and the classical pressure become equally important at $\lambda^*$ when
\be
m\Phi = \frac{M|\lambda^*|}{4m^3} \psi^2 
\ee
Using $\Phi \sim \frac{GM}{R}$ and the radial size $R\sim(GMm^2)^{-1}$ this implies a critical coupling of $\lambda^* \sim -G m^4 R^2 \sim -\frac{M_{pl}^2}{M^2}$. For a given coupling $\lambda$, the non-gravitational limit applies for stars with $R< R^*$ where the critical radius is $R^*=\frac{M_{pl}\sqrt{|\lambda|}}{m^2}$ with a corresponding critical mass $M^*= (Gm^2 R^*)^{-1} = M_{pl} |\lambda|^{-\frac{1}{2}}$ . 

For $\lambda<\lambda^*$ we can rewrite Eq.~\ref{gpp-seq} as
\be
\frac{1}{2m}\nabla^2  \left(\frac{\nabla^2 \psi}{\psi}\right) = \frac{M|\lambda|}{4m^3} \nabla^2 \psi^2. 
\ee
Using the scaling behavior, we see that the radial size of the star in the non-gravitational limit is $R \sim \frac{M |\lambda|}{m^2}$. The radius increases linearly with the mass of the star. Since at larger radius $R>R^*$ we approach the non-interacting limit in which the mass decreases for increasing radius, we find that there is a maximum possible mass for boson stars $M^{max}= M^* \sim M_{pl} |\lambda|^{-\frac{1}{2}}$. However, it has be shown in \cite{Chavanis:2011zi}, that the solutions for $R<R^*$ are unstable with respect to perturbations. Therefore boson stars in the non-gravitational limit cannot be realized in nature, at least for this simple class of interactions. For more complicated interactions, such stars can exist, and fall under the general class of Q-balls \cite{Kusenko:1997zq, Kusenko:1997yj, Kleihaus:2005me, Hartmann:2012gw}.

\subsection{Scaling Invariance}

Following the terminology of \cite{Kling:2017mif} we introduce the dimensionless variables 
\be
 V&=\frac{e}{2m}-\frac{\Phi}{2}, \quad &S&=\sqrt{\frac{\pi G N}{2m}} \psi, \\
 x&=2rm,  \quad
 &\Lambda&=\frac{\lambda}{4 \pi G m^2}.
\label{gpp-dimles-vars}
\ee
We can then rewrite the Gross-Pitaevskii-Poisson equations given in Eq.~\ref{gpp-seq} and \ref{gpp-neq} as
\be
\nabla^2 V = -S^2 \quad \text{and} \quad \nabla^2 S = - VS + \Lambda S^3.
\label{gpp-dimensionless}
\ee
The wave function normalization condition $\int \psi^2 dV =1$ becomes 
\be
\int_0^\infty x^2 S^2 dx = GMm
\label{gpp-mass}
\ee
where $M=Nm$ is the mass of the boson star. 

Let us note that Eq.~\ref{gpp-dimensionless} and \ref{gpp-mass} are invariant under the scaling 
\be
\hspace*{-0.4cm}
x \to \frac{x}{f},  \ S \to f^2 S, \ V \to f^2 V, \ M \to f M, \ \Lambda \to \frac{\Lambda}{f^2}
\label{gpp-scaling}
\ee 
where $f$ is a scaling factor. This implies that we can relate different solutions of the Gross-Pitaevskii-Poisson equations corresponding to different boson star masses $M$ and couplings $\Lambda$ through rescaling. We will make use of this scale invariance and solve for the Gross-Pitaevskii-Poisson equations at a fixed reference scale $k$. A particularly useful choice for our discussion is to set $-k^2 = V(\infty)=\frac{e}{2m}$ which transforms as $k \to f k$. We can then introduce the scale invariant coordinate $z$, wave function $s$, potential $v$, mass $\beta$ and coupling $\gamma$ via
\be
z &= kx, \quad S = k^2 s, \quad V = k^2 v, \\
\gamma &= k^2 \Lambda, \quad\quad   GMm=2k\beta.
\label{gpp-rescale}
\ee
Using the scale independent variables, we can write the Gross-Pitaevskii-Poisson equations as
\be
\nabla^2 s = - sv + \gamma s^3 \;\;\; \text{and}\;\; \nabla^2 v = -s^2.
\label{gpp-scaleinvariantgpp}
\ee  
The scale choice implies the boundary condition $v(\infty)=-1$. The solution corresponding to a boson star with mass $M$ can be obtained by performing the rescaling given in Eq.~\ref{gpp-rescale} with $k =\frac{GMm}{2\beta}$. In the following section, we will obtain an approximate analytical form for $s,v$. 

\section{Weakly Coupled Systems and Series Expansion}

\subsection{Series Expansion}
\label{sec-series}

We have seen in the discussion of the non-self-interacting case, that we can describe the profile of the boson star through an infinite series for the wave function and potential. 

Following the same approach as for the non-self-interacting case \cite{Kling:2017mif}, we will describe the profile at both small and large radii through a series expansion of the wave function and potential. At small radii, the profile can be described via an (even) polynomial around the center if the boson star $z=0$
\be
s = \sum_{n=0}^{\infty} s_n z^n \;\;\text{and}\;\; v = \sum_{n=0}^{\infty} v_n z^n \ .
\label{smallz-expansion}
\ee
Eq. \ref{gpp-scaleinvariantgpp} then leads to the recursion relations
\be
-\sum_{m=0}^{n} &s_m v_{n-m} +\gamma\sum_{m=0}^{n}\sum_{\ell=0}^{m} s_\ell s_{m-\ell} s_{n-m}\\
&= (n+2)(n+3)  s_{n+2} \\
-\sum_{m=0}^{n} &s_m s_{n-m}=(n+2)(n+3)  v_{n+2}  .
\ee
The smoothness of the profile at the origin implies $s_1=v_1=0$, and therefore, also all odd coefficients $s_{2n+1} , v_{2n+1}$ vanish. The profile at small radius $z$ can therefore be fully parametrized in terms of the wave function and potential at the origin: $s_0$ and $v_0$.

At large radius, we can expand the profile using the series expansion
\be
\hspace{-0.2cm}
s =\hspace{-0.3cm} \sum_{n,m=0,0}^{\infty,\infty}\hspace{-0.3cm} s^n_m 
\left( \frac{e^{-z}}{z^\sigma} \right)^n \hspace{-0.2cm} z^{-m}  , \; \; \; 
v =\hspace{-0.3cm} \sum_{n,m=0,0}^{\infty,\infty}\hspace{-0.3cm} v^n_m 
\left( \frac{e^{-z}}{z^\sigma} \right)^n \hspace{-0.2cm} z^{-m} .
\label{largez-expansion}
\ee
By matching the coefficients, Eq.~\ref{gpp-scaleinvariantgpp} we obtain the recursion relations
\be
 \hspace{-0.4cm} \sum_{p,q=0,0}^{n,m} &s^p_q v^{n-p}_{m-q} + n^2 \, s^n_m  
+ 2 n(n\sigma+m-2) \; s^n_{m-1}  \\
&+  (n\sigma+m-2)(n\sigma+m-3)\, s^n_{m-2} \\
&\quad= \gamma \sum_{p,q=0,0}^{n,m} \sum_{r,t=0,0}^{p,q} s^{r}_{t}\, s^{p-r}_{q-t}\, s^{n-p}_{m-q}
\label{largez-eqs}
\ee
\vspace{-0.5cm}
\be
\hspace{-0.4cm} \sum_{p,q=0,0}^{n,m} &s^p_q s^{n-p}_{m-q} + n^2 \, v^n_m  
+ 2 n(n\sigma+m-2) \; v^n_{m-1}  \\
&+ (n\sigma+m-2)(n\sigma+m-3)\,v^n_{m-2} 
= 0. 
\label{largez-eqv}
\ee
Let us note the following properties of the solution: 
i) Normalizability requires $s^0_0=0$. Eq.~\ref{largez-eqs} then implies that all coefficients $s^0_m$ vanish as well. This means that the wave function decays at least exponentially. 
ii) Eq.~\ref{largez-eqv} then implies that all $v^0_m=0$ for $m>1$. This means that at large radius, the potential is described by the Newtonian potential $v^{(0)}= v^0_0+\frac{v^0_1}{z}=-1+\frac{2\beta}{z}$. Here we both used the boundary condition imposed by our scale choice, $v_0^0=-1$, and used the notation introduced in Eq.~\ref{gpp-rescale}, $v_1^0=2\beta$. All other terms in the expansion of $v$ are at least exponentially suppressed. 
iii) Setting $n=m=1$, Eq.~\ref{largez-eqs} can be written as $2\beta=v_1^0=2(1-\sigma)$. This is a remarkable result: the exponent $\sigma$ in the series expansion is related to the the total mass of the system $\sigma = 1-\beta$.
iv) As derived in \cite{Kling:2017mif}, the leading order, $n=1$, solution for the wave function at large radius is given by the Whittaker function
\be
s^{(1)}= \frac{ \alpha}{2^{\beta}z} W_{\beta,-\frac{1}{2}}(2z).
\label{series-whittaker}
\ee
Here we have introduced the normalization parameter $\alpha=s_0^1$. 
v) The large radius solution can be fully parametrized by the expansion parameters $\alpha=s_0^1$ and $\beta=v_1^0$. The remaining coefficients can then be computed using Eq.~\ref{largez-eqs} and \ref{largez-eqv}. Note however, that the series expansion in Eq.~\ref{largez-expansion} does only converge for $m<M$, where $M$ is finite.
vi) Eq.~\ref{largez-eqs} and \ref{largez-eqv} further imply that the potential contains only non-vanishing components $v^n_m$ for even $n$ while the wave function only has non-vanishing components $s^n_m$ for odd $n$. 

For practical purposes, we will truncate the infinite series in Eq.~\ref{smallz-expansion} and \ref{largez-expansion} and only take into account the leading terms with $n\leq N$ and $m \leq M$. Let us define the truncated series expansion at small and large radius $z$ as
\be
s_{(N)} = \hspace{-0.1cm} \sum_{n=0}^{N} s_n z^n \ \  \text{and} \ \ 
s_{(M)}^{(N)} =\hspace{-0.3cm} \sum_{n,m=0,0}^{N,M}\hspace{-0.3cm} s^n_m 
\left( \frac{e^{-z}}{z^\sigma} \right)^n \hspace{-0.2cm} z^{-m}   .
\label{series-truncated}
\ee

\subsection{Expansion Parameters } 

\begin{figure*}[thb]
\centering
\includegraphics[width=0.45\textwidth]{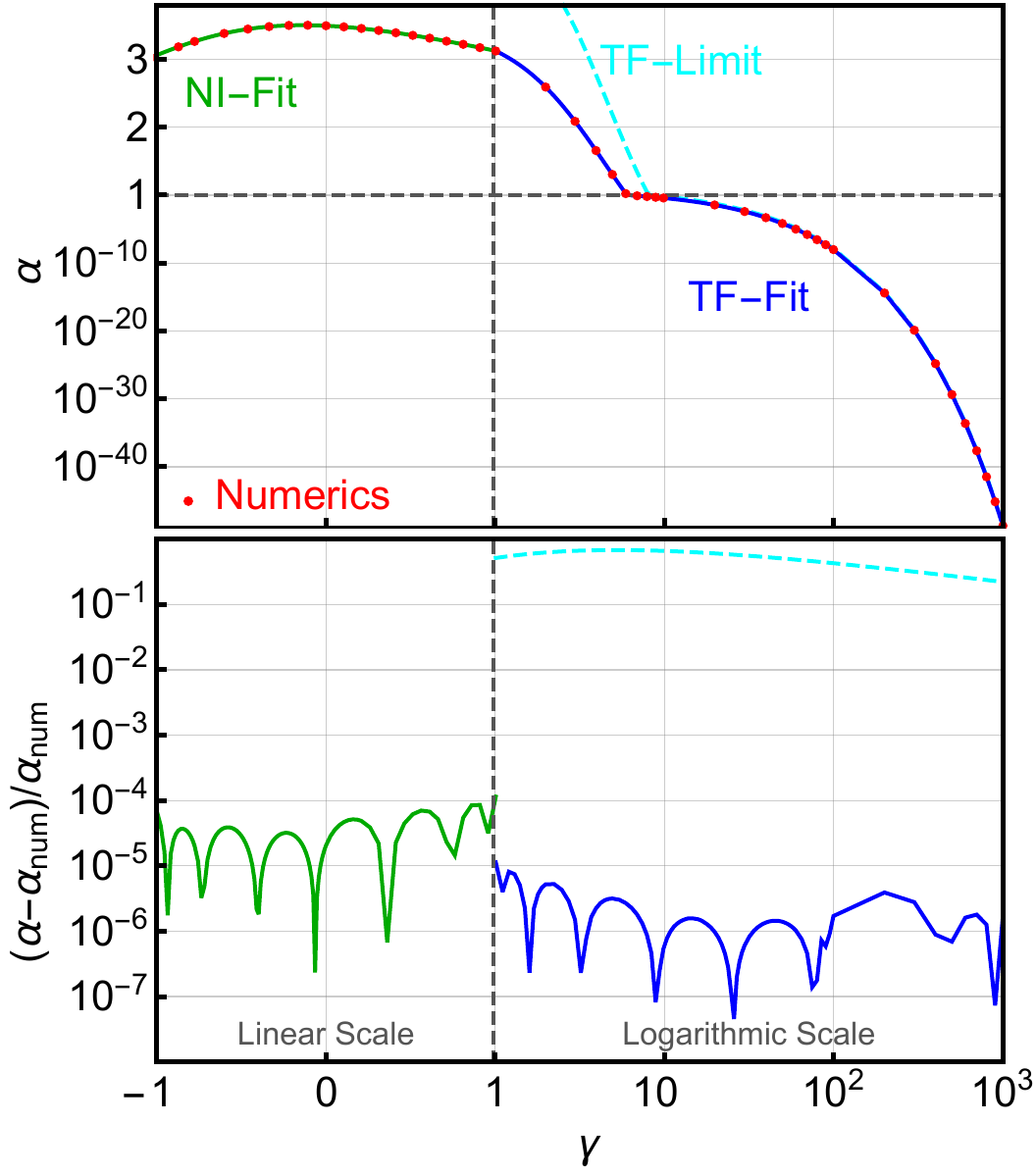}
\includegraphics[width=0.45\textwidth]{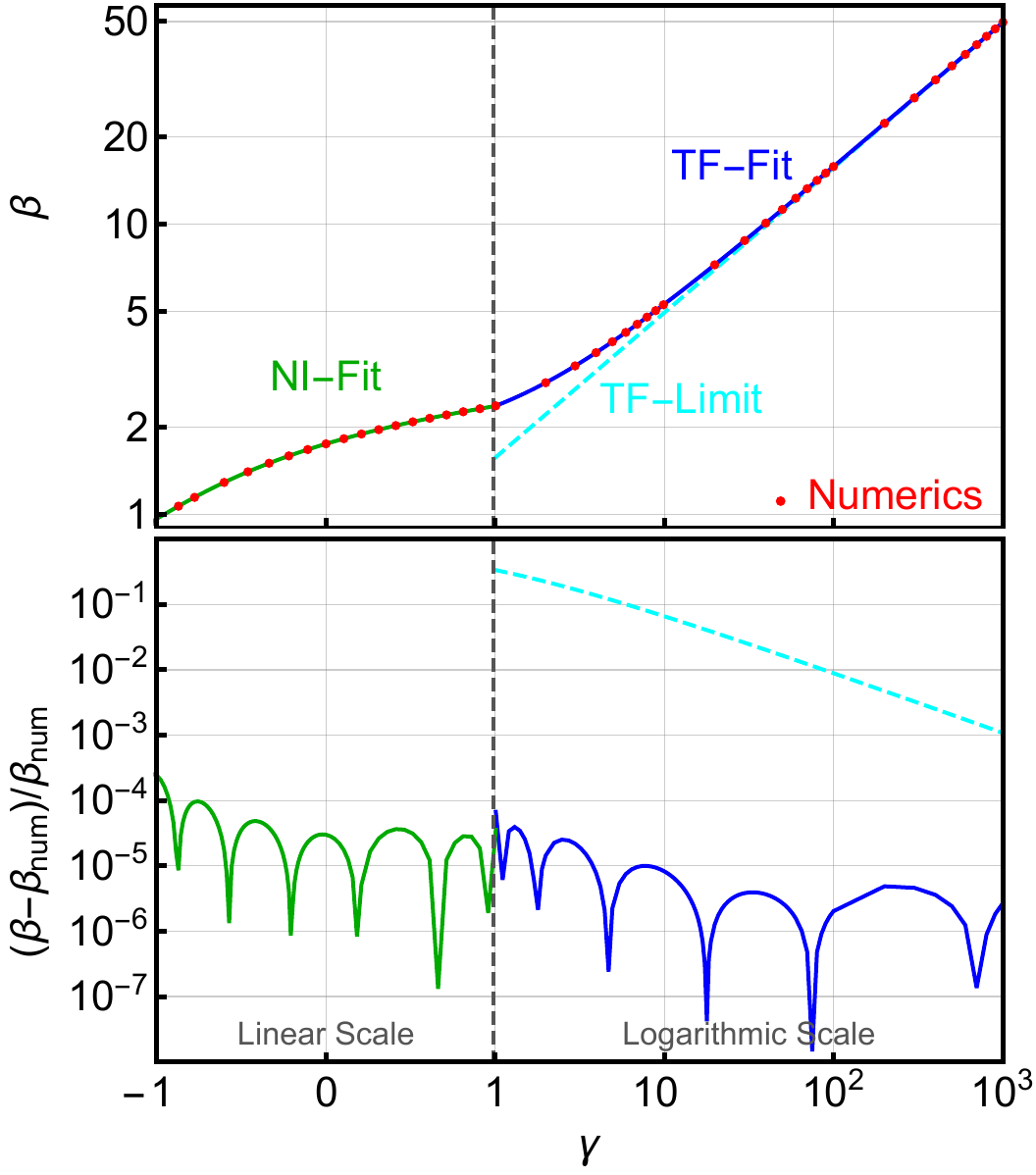}\\
\vspace{0.0cm}
\includegraphics[width=0.45\textwidth]{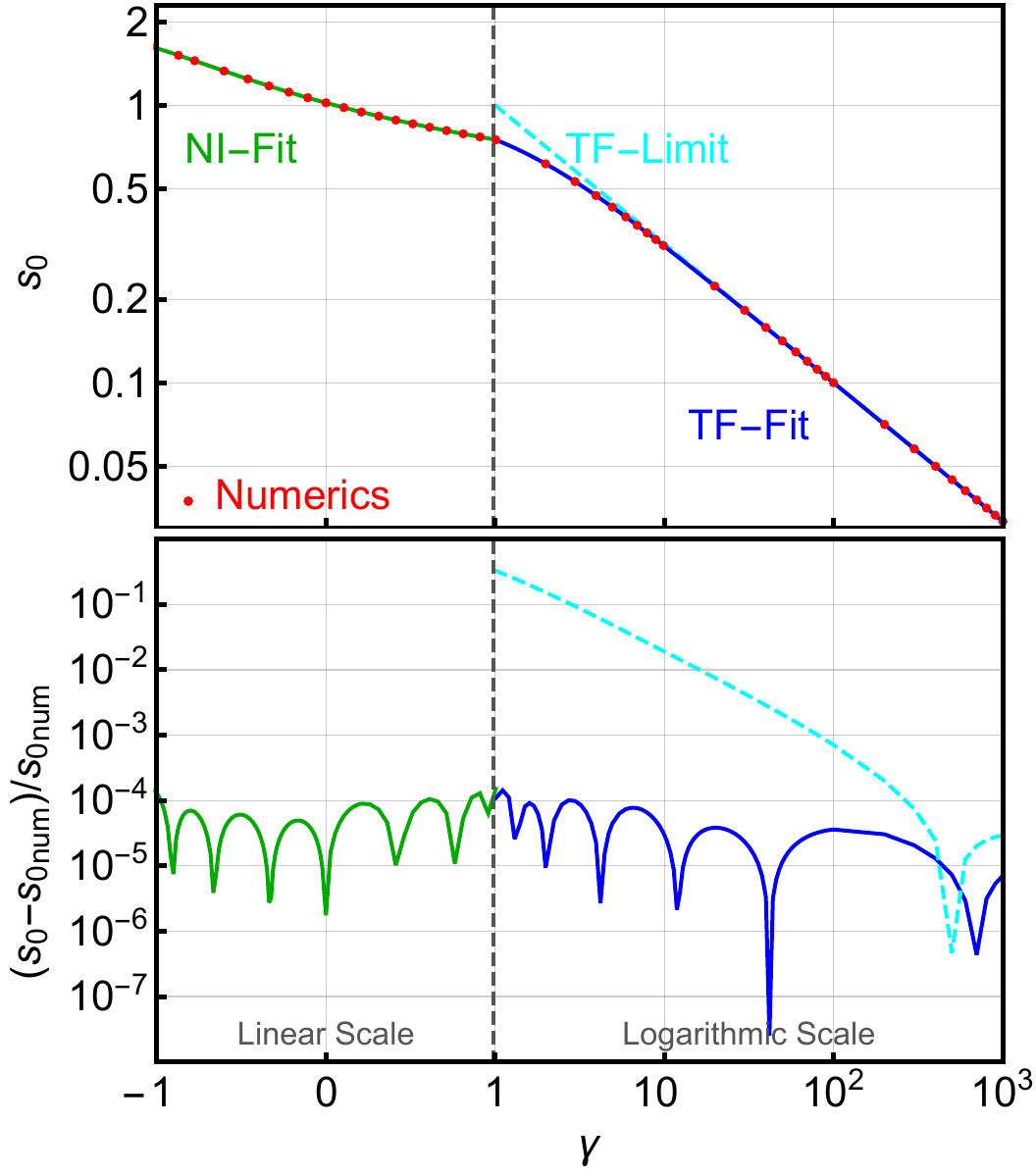}
\includegraphics[width=0.45\textwidth]{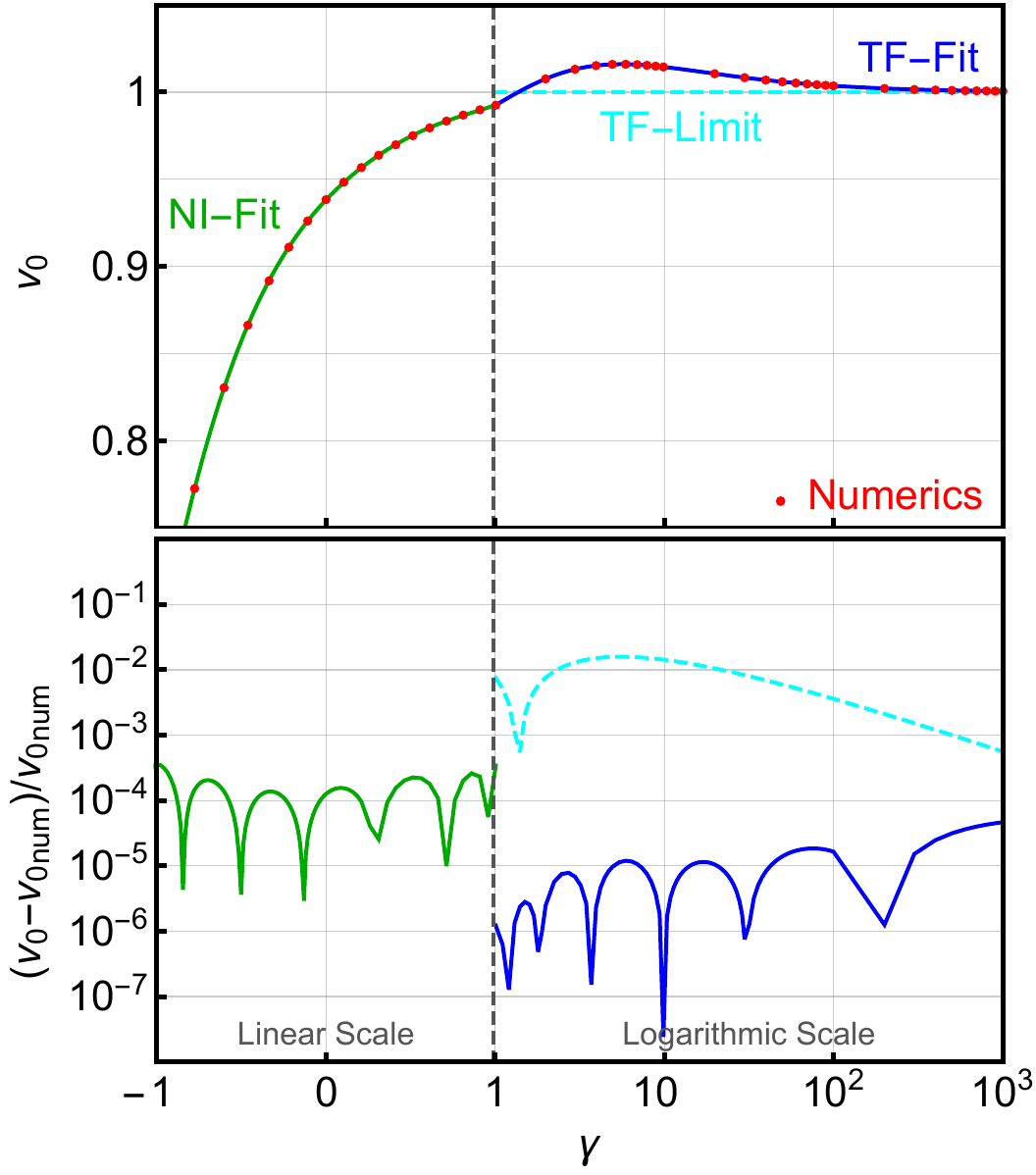}
\vspace{0.0cm}
\caption{The upper panels show the expansion parameters $\alpha$ (upper left), $\beta$ (upper right), $s_0$ (lower left) and $v_0$ (lower right) as function of the coupling parameter $\gamma$. The numerical solution is shown as red dots in the upper panels. The analytic fits to the numerical results are shown in green for the Non-Interacting regime as discussed in Sec.~\ref{sec:ni-fit} and in blue for the Thomas-Fermi regime as discussed in Sec.~\ref{sec:tf-fit}. The results corresponding to the Thomas-Fermi limit discussed in Sec.~\ref{sec:tf-limit} are shown in cyan for comparison. The lower panels show the accuracy of the analytic solutions with respect to the numerical solution. 
}
\label{series:param-fit}
\end{figure*}

As we have seen in the previous section, the series expansion for $s$ and $v$ at small and large radius can be fully parametrized by the four expansion parameters $\alpha,\beta,s_0,v_0$. In the non-interacting case $\gamma=0$ these were just numbers, while in the general case they will be functions of the coupling $\gamma$
\be
\alpha(\gamma), \;\; \beta(\gamma), \;\; s_0(\gamma), \;\; v_0(\gamma).
\ee

Following the strategy from \cite{Kling:2017mif}, we first obtain the expansion parameters from numerical simulations. In a second step, we provide an analytic form for the expansion parameter. 

To obtain a numerical solution of the boson star's profile, it is convenient to solve the Gross-Pitaevskii-Poisson equations as given in Eq.~\ref{gpp-dimensionless} using the boundary condition $V(0)=1$. The authors of \cite{Tod1992} have shown that the solutions of Eq.~\ref{gpp-dimensionless} can then be parametrized by the central value of the wave function, $S_0=S(0)$ and categorized into three distinct classes: for $S_0>S_0^*$ the wave function diverges for at large radius towards positive infinity, for $S_0=S_0^*$ the wave function converges to zero, is positive definite and square integrable, while for $S_0<S_0^*$ the wave function diverges at large radius towards negative infinity. 

Using a Runge-Kutta 4 method with constant step size $\Delta_x$, we perform the numerical integration until the wave function starts to diverge and iteratively optimize the central value of the wave function $S_0$ to find $S_0^*$. The precision of the wave function needed for the numerical solution to stay finite until a large value of $x$, which is needed to fit the large range solution, increases exponentially with the radial coordinate $x$.  The accuracy of the numerical solution is limited by the step size $\Delta_x$. In this study we use a precision of up to 150 significant figures and $\Delta_x = 10^{-3}$, providing an accuracy of the solution of order $\mathcal{O}\left(\Delta_x^4 \right) \approx 10^{-12}$. 

The Whittaker solution parametrization given in Eq.~\ref{series-whittaker} always describes the wave function profile at large radius when the density is small enough that the self-interaction becomes negligible. However, the central mass parameter $\beta$ and the normalization $\alpha$ depend on the central profile of the star and therefore the coupling parameter $\gamma$.

To obtain the expansion parameters $\alpha,\beta$ as well as the scaling parameter $k$ in Eq.~\ref{gpp-rescale}, we fit the leading order profiles, the Whittaker solution $S(x) = \frac{ k \alpha}{2^\beta x} W_{\beta,-\frac{1}{2}}(2kx)$ and the Newtonian potential $V(x) = -k^2 + \frac{2k\beta}{x}$, to the numerical solution for $V$ and $S$ at large $x$. To avoid systematic effects due to the truncation of subleading terms $n>1$ of the series expansion in Eq.~\ref{largez-expansion}, we restrict the fitting range to $x>x^*$, where the fraction of mass outside radius $x^*$ contributed less than $10^{-12}$ to the total mass of the boson star. The expansion parameters at small radius and the coupling are obtained via $s_0=k^{-2} S_0$, $v_0 = k^{-2}$ and $\gamma=k^2 \Lambda$. 

In Fig.~\ref{series:param-fit} we show the dependence of the expansion parameters $\alpha$ (upper left), $\beta$ (upper right), $s_0$ (lower left) and $v_0$ (lower right) as function the coupling parameter $\gamma$ as red dots. Note that the horizontal axis switches from a linear scale to a logarithmic scale at $\gamma=1$ as indicated by the dashed black line. This indicated the transition between the weakly coupled regime $\gamma<1$ and the strongly coupled regime $\gamma>1$. 

\subsection{Fit for Weak Couplings: $-1<\gamma<1$}
\label{sec:ni-fit}

A vanishing self-coupling $\gamma=0$ indicates the non-interacting limit. The corresponding results for the expansion parameters are presented in \cite{Kling:2017mif}. If the self-coupling is weak, $|\gamma| < 1$, we can treat the classical pressure as perturbation. In this case we can write the the expansion parameters as a series expansion in the coupling $\gamma$ around the non-interacting solution. To obtain the coefficients of this expansion, we fit a 6th-degree polynomial to the numerical solutions. We can write the result in an analytic form as 
\begin{widetext}
\be
\alpha(\gamma)& 
=3.495059
 -0.117682\,\gamma
 -0.391600\,\gamma^2
+0.191882\,\gamma^3
 -0.041828\,\gamma^4
 -0.041507\,\gamma^5
+0.033020\,\gamma^6\\
\beta(\gamma)& 
=1.752717
+0.703934\,\gamma
 -0.109101\,\gamma^2
+0.013436\,\gamma^3
+0.017778\,\gamma^4
 -0.018281\,\gamma^5
+0.005129\,\gamma^6\\
s_0(\gamma)&
=1.021494
 -0.390946\,\gamma
+0.171489\,\gamma^2
 -0.064820\,\gamma^3
+0.004328\,\gamma^4
+0.028849\,\gamma^5
 -0.017732\,\gamma^6\\
v_0(\gamma)&
=0.938204
+0.102743\,\gamma
 -0.080310\,\gamma^2
+0.058708\,\gamma^3
 -0.037703\,\gamma^4
 -0.002557\,\gamma^5
+0.013512\,\gamma^6 \, .
\label{analytic-NI}
\ee
\end{widetext}

The result is shown in Fig.~\ref{series:param-fit} as a green line. The lower panels indicate the accuracy of the analytic form with respect to the numerical solution. We can see that for all four expansion parameters the solution from Eq.~\ref{analytic-NI} reproduces the numerical results with accuracy better of $\mathcal{O}(10^{-4})$ in the range $-1 < \gamma < 1$. 

As mentioned before, boson stars with attractive self-coupling and small radius $R$, or equivalently large negative coupling $\gamma$, become unstable with respect to perturbations. We show in Sec.~\ref{AppUsing}, that this happens at $\gamma_{min} = -0.722$. The solutions for $\gamma<\gamma_{min}$ are unphysical. 

\section{Strongly Coupled Systems}
\label{sec-tfl}

\subsection{Thomas Fermi Limit}
\label{sec:tf-limit}

In the previous section we have discussed the weakly coupled scalar field. Let us now consider the case of a large repulsive self-coupling $\gamma \gg 1$. In this case the quantum pressure becomes negligible and the classical pressure balances gravity. This scenario is known as Thomas-Fermi limit and has been examined in \cite{2007JCAP...06..025B,Slepian:2011ev} We can write the Gross-Pitaevskii-Poisson equations as 
\be
\gamma \nabla^2 v + v =0  \; \;  \text{and} \quad v =  \gamma s^2. 
\label{tfl-eq}
\ee
The (normalizable) solution for the profile is given by 
\be
v= v_0 \;  \text{sinc}(z/\sqrt{\gamma})
\quad  \text{and}  \quad
s = s_0 \left[ \text{sinc}(z/\sqrt{\gamma}) \right]^\frac{1}{2}
\label{ft-profile}
\ee
where $\text{sinc}(x)=\sin(x)/x$. The wave function becomes zero at $Z = \pi \sqrt{\gamma}$, implying that the boson star is compact and has a radius $Z$. At $z>Z$ the wave function remains zero and the potential is described by the Newtonian potential $v(z)= -1+\frac{2\beta}{z}$. 

Using that exterior and interior solution for the potential of the star have to match at surface, $v(Z)=0$, we can solve for the mass parameter $\beta = \frac{1}{2}Z = \frac{\pi}{2} \sqrt{\gamma}$. We can further use the normalization condition (see Eq.~\ref{gpp-mass}) 
\be
\hspace{-0.1cm} 
2 \beta = \int s^2 z^2 dz 
= s_0^2 \int_0^{\pi \sqrt{\gamma}} \hspace{-0.3cm} \text{sinc}(z/\sqrt{\gamma}) z^2 dz  
= s_0^2 \pi \gamma^{\frac{3}{2}}
\ee
to obtain the central density $s_0 = \gamma^{-\frac{1}{2}}$. Using Eq.~\ref{tfl-eq}, this implies that the central value for the potential is $v_0=1$. 

\bigskip 

So far, we have ignored the effects of quantum pressure to the boson star's profile. However, close to the star's radius $Z$, the density drops until eventually the self-interaction becomes negligible. The outer part of the star can therefore be described by the non-interacting solution, which at leading order can be written in terms of the Whittaker function. In the following, we estimate the remaining expansion parameter $\alpha$ by matching the Thomas-Fermi solution for the interior of the star with the Whittaker solution for the exterior part. 

We define the matching point $z^*$ as the radius where the quantum and classical pressure terms become equal, $\nabla^2 s(z^*) = \gamma s(z^*)^3 $. Using the Thomas-Fermi profile from Eq.~\ref{ft-profile}, we can write this condition as 
\be
\frac{\gamma s_0^4}{ 8 z^4 s^3} &\Big[ 1 + 3 \frac{z^{*2}}{\gamma} - \left(1 + \frac{z^{*2}}{\gamma} \right) \cos\left(\frac{2z^*}{\sqrt{\gamma}}  \right)\\
& - \frac{2z^*}{\sqrt{\gamma}}  \sin\left(\frac{2z^*}{\sqrt{\gamma}}\right) \Big] 
= 
\frac{s_0^3 \gamma^\frac{3}{2}}{z^{*3}}     \sin\left(\frac{z^*}{\sqrt{\gamma}}  \right)^3
\ee
We can write the matching point as $z^*=(\pi-\delta)\sqrt{\gamma}$ and expand the matching condition in $\delta$. Keeping only the linear terms and using $s_0 = \gamma^{-\frac{1}{2}}$, we can solve for $\delta$ and obtain $\delta = (4 \gamma / \pi)^{-\frac{1}{3}}$. The wave function at $z^*$ has a value 
\be
s(z^*) 
=  s_0\,  \text{sinc}(\pi-\delta)^\frac{1}{2}  
\approx  s_0 \Big(\frac{\delta }{\pi}\Big)^\frac{1}{2} 
= (2\pi\gamma^2)^{-\frac{1}{3}}.
\label{szstar}
\ee

Following \cite{buchholz1969confluent} (see Sec.~8, Eq.~18) we can write the Whittaker function $W_{\beta,-\frac{1}{2}}(2z)$ for $z = 2 \beta = \pi \sqrt{\gamma} $ as
\be
\hspace{-0.2cm}
W_{\beta,-\frac{1}{2}}(2z) \approx \Gamma\Big(\frac{1}{3}\Big) \Big(\frac{2z}{6 \pi^3} \Big)^\frac{1}{6}\hspace{-0.2cm}
\exp\left[\beta \log\Big(\frac{\beta}{e}\Big) +\frac{1}{12\beta}\right]
\ee
Using the form of the exterior profile as given in Eq.~\ref{series-whittaker} and matching it to the Thomas-Fermi solution at $z^*$ in Eq.~\ref{szstar} allows us to extract the expansion parameter $\alpha$. We find 
\be
\hspace{-0.3cm}
\alpha 
&= \frac{2^\beta z^* s(z^*)}{W_{\beta,-\frac{1}{2}}(2z^*)} 
\approx 
\frac{\big(\frac{3}{4}\big)^\frac{1}{6} \Gamma\big(\frac{1}{3}\big)^{-1}  \pi \gamma^{-\frac{1}{4}} }{
\exp\hspace{-0.1cm}\Big[\frac{\pi}{2} \gamma^\frac{1}{2} \log\big(\frac{\pi}{4e} \gamma^\frac{1}{2} \big) +\frac{1}{6 \pi} \gamma^{-\frac{1}{2}} \Big]}\hspace{-0.5cm}
\label{tf-alpha}
\ee
We have already seen that in the Thomas-Fermi limit the remaining expansion parameters are given by
\be
\beta = \frac{\pi}{2} \gamma^\frac{1}{2}, \quad s_0 =  \gamma^{-\frac{1}{2}}, \quad v_0 = 1\ .
\label{tf-beta}
\ee
We have therefore obtained a simple analytic form for all expansion parameters in the limit of large self-couplings. The results from Eq.~\ref{tf-alpha} and \ref{tf-beta} are shown in Fig.~\ref{series:param-fit} as dashed cyan lines.  

Note that for increasing self-coupling $\gamma$, the expansion parameter $\alpha$ exponentially decreases leading to a sharp drop of the wave function profile at $z=Z$. The bosons become more and more confined in the inner part while the large radius tails of the wave function vanish. 

\subsection{Fit for Strong Couplings: $\gamma>1$}
\label{sec:tf-fit}

In the above discussion of the Thomas-Fermi limit, we have neglected the effect of the quantum pressure on the structure of the star. For large but finite self-couplings $\gamma$, we can consider the quantum pressure as a perturbation to the Thomas-Fermi limit. We include this perturbation as a correction factor to the Thomas-Fermi solution, which we can express asseries expansion in the (inverse) coupling parameter. The coefficients of this expansion are obtained from a fit to the numerical solution. We can write the result in an analytic form as
\begin{widetext}
\begin{align}
\hspace{-0.3cm}
\alpha(\gamma)&= \alpha^{TF}
\big[
0.603380
+0.485970 \gamma^{-\frac{1}{6}} \hspace{-0.05cm}
 -4.422475 \gamma^{-\frac{2}{6}}\hspace{-0.05cm}
+8.719758 \gamma^{-\frac{3}{6}}\hspace{-0.05cm}
 -8.363927 \gamma^{-\frac{4}{6}}\hspace{-0.05cm}
+4.397913 \gamma^{-\frac{5}{6}}\hspace{-0.05cm}
 -1.001027 \gamma^{-1}
\big] \nonumber \\
\hspace{-0.3cm}
\beta(\gamma)&= \makebox[0pt][l]{$\beta^{TF}$}\phantom{\alpha^{TF}} 
\big[
1
 -0.001478 \, \gamma^{-\frac{1}{3}}
+0.045642 \, \gamma^{-\frac{2}{3}}
+0.823049 \, \gamma^{-1}
 -0.590994 \, \gamma^{-\frac{4}{3}}
+0.347840 \, \gamma^{-\frac{5}{3}}
 -0.118132 \, \gamma^{-2}
\big] \nonumber\\
\label{analytic-TF}
\hspace{-0.3cm}
s_0(\gamma)&=  \makebox[0pt][l]{$s_0^{TF} $}\phantom{\alpha^{TF}} 
\big[
1
+0.003712\, \gamma^{-\frac{1}{2}}
 -0.067139\, \gamma^{-1} 
 -0.436976\, \gamma^{-\frac{3}{2}}
 -0.107433\, \gamma^{-2} \,
+0.687868\, \gamma^{-\frac{5}{2}}
 -0.327405\, \gamma^{-3}
\big] \\
\hspace{-0.3cm}
v_0(\gamma)&=  \makebox[0pt][l]{$v_0^{TF} $}\phantom{\alpha^{TF}}  
\big[
1
+0.008062\, \gamma^{-\frac{1}{2}}
+0.388054\, \gamma^{-1}
 -1.245466\, \gamma^{-\frac{3}{2}}
+1.486280\, \gamma^{-2}\,
 -0.823199\, \gamma^{-\frac{5}{2}}
+0.178605\, \gamma^{-3}
\big] \nonumber
\end{align}
with the leading order Thomas-Fermi solutions 
\begin{align}
\alpha^{TF} =
\big(\frac{3}{4}\big)^\frac{1}{6} \Gamma\big(\frac{1}{3}\big)^{-1}  \pi \gamma^{-\frac{1}{4}} 
\exp\Big[-\frac{\pi}{2} \gamma^\frac{1}{2} \log\big(\frac{\pi}{4e} \gamma^\frac{1}{2} \big) -\frac{1}{6 \pi} \gamma^{-\frac{1}{2}} \Big] , 
\quad  \beta^{TF} = \frac{\pi}{2} \gamma^\frac{1}{2}, 
\quad s_0^{TF}  =  \gamma^{-\frac{1}{2}}, 
\quad v_0^{TF}  = 1 \ .
\end{align}
\end{widetext}
Note that the correction factor for the expansion parameter $\alpha$ contains a non-unity constant term. This should not be surprising, considering that we obtained $\alpha$ in the Thomas-Fermi limit by matching the Thomas-Fermi solution and the Whittaker-solution at a matching point $z^{*}$, even though both solutions describe the profile only poorly at this point. 

The results are shown in Fig.~\ref{series:param-fit} as a blue line. We can see that for all four expansion parameters the solution from Eq.~\ref{analytic-TF} reproduces the numerical results with accuracy of  $\mathcal{O}(10^{-5})$ for self-couplings $\gamma > 1$. 

\begin{figure}[t h b ]
\centering
\includegraphics[width=0.45\textwidth]{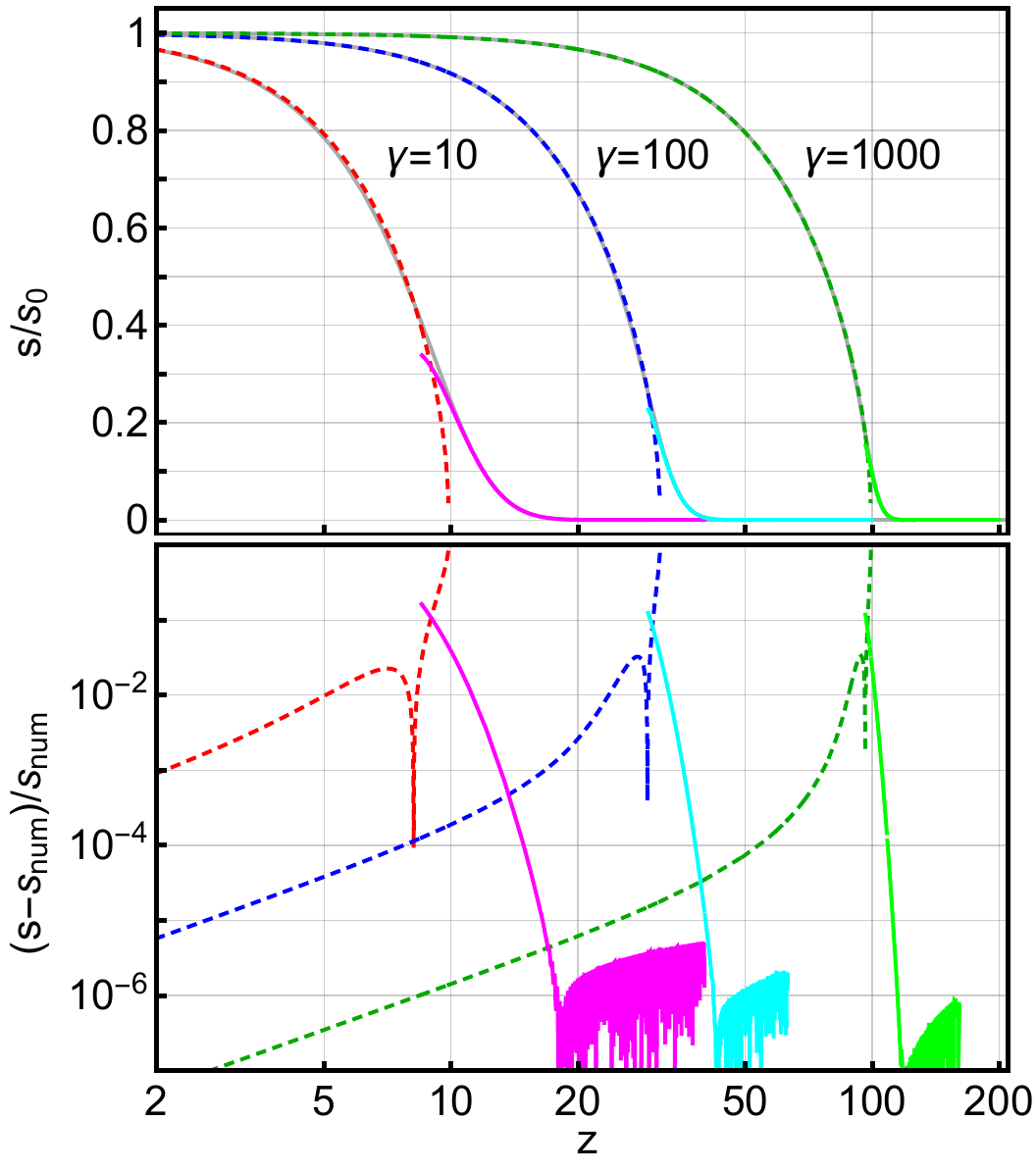}
\caption{The upper panel shows the wave function profile $s(z)$ for the couplings $\gamma=10$ (left), $100$ (center) and $1000$ (right). We show the numerical solution (gray), the Thomas-Fermi approximation (dashed) for the central profile (see Eq.~\ref{ft-profile}) and the Whittaker approximation (solid) for the outer profile (see Eq.~\ref{series-whittaker}). The lower panel shows the accuracy of the approximate solutions with respect to the numerical solution, $(s-s_{\text{num}})/s_{\text{num}}$, as indicated by the colored dashed lines. We use the expansion parameters according to Eq.~\ref{analytic-TF}. 
}
\label{tf:limit}
\end{figure}

Using the analytic expression for the expansion parameters, we can now obtain the profile of the boson star for a given value of the self-coupling $\gamma$. For very large self-couplings, $\gamma \gtrsim 10$, we can directly use the Thomas-Fermi and Whittaker solution to describe the central and outer profile of the star. This is shown in Fig.~\ref{tf:limit} for $\gamma=10$ (red), $100$ (blue) and $1000$ (green). The upper panel shows the wave function profile $s(z)$ normalized by its central value $s_0$. The numerical solution is shown in gray, the Thomas-Fermi solution for the central profile given by Eq.~\ref{ft-profile} as dashed lines and the Whittaker solution given by Eq.~\ref{series-whittaker} as solid lines. The lower panel shows the accuracy of the approximate solutions with respect to the numerical solutions. 

We can see that for large couplings, a combination of the Thomas-Fermi and Whittaker solutions describe profile with an accuracy of $\mathcal{O}(10^{-6})$ in the center and outside the star, corresponding to the accuracy of the expansion parameters. However, even for large couplings, there remains a surface region around $z \approx Z$ that is poorly described by both the Thomas-Fermi and Whittaker solutions. Even though the wave function, and therefore the mass density, is small in the surface region, the fraction of the star's mass contained in it can still be sizeable.

\begin{figure}[t h b]
\centering
\includegraphics[width=0.45\textwidth]{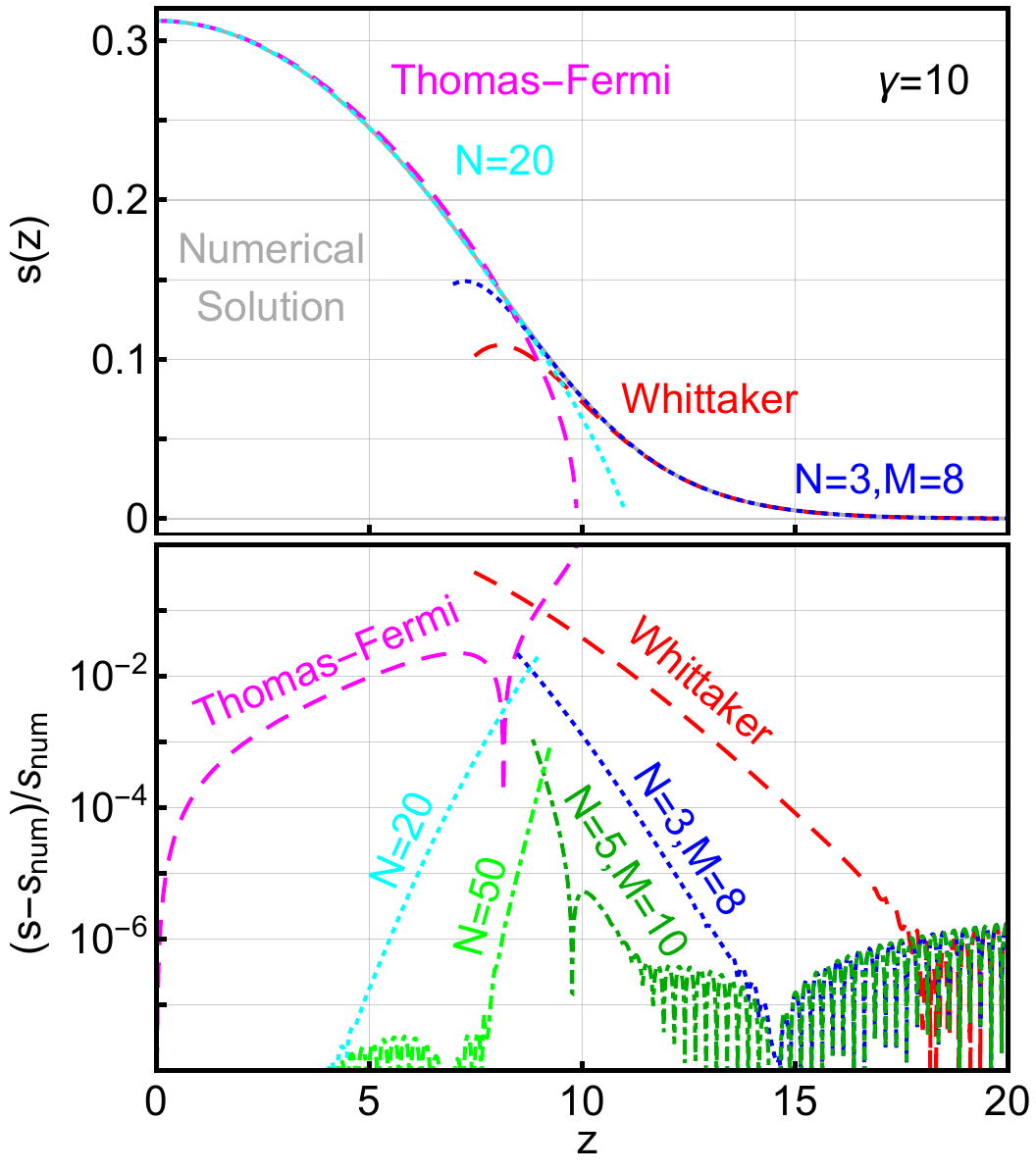}
\caption{The upper panel shows the numerical solution (gray), the Thomas-Fermi approximation as given in Eq.~\ref{ft-profile} (magenta dashed), the Whittaker approximation as given in Eq.~\ref{series-whittaker} (red dashed), the truncated series expansion of the wave function at small radius with $N=20$ (cyan dotted) and with at large radius with $N=3, \ M=8$ (blue dotted) as given in Eq.~\ref{series-truncated}. The lower panel shows the accuracy of the approximate solutions with respect to the numerical solution, $(s-s_{\text{num}})/s_{\text{num}}$. We use the expansion parameters according to Eq.~\ref{analytic-TF}. 
}
\label{tf:solution}
\end{figure}

To obtain a better description, in particular for the surface region, we can use the series expansion derived in Sec.~\ref{sec-series}. This is shown in Fig.~\ref{tf:solution} for an intermediate sized self-coupling $\gamma=10$. The upper panel shows the wave function profile $s(z)$. Besides the numerical solution (gray), the Thomas-Fermi solution (magenta) and the Whittaker solution (red), we show the truncated series expansion as given in Eq.~\ref{series-truncated} with $N=20$ for the inner solution (cyan) and $N=3,\ M=8$ for the outer solution (blue). The corresponding accuracy of the solution is shown in the lower panel. We can see that including higher order terms in the truncated series expansion increases the accuracy of the solution by one order of magnitude in the surface region and up to four orders of magnitude in the center and outside star. For comparison we also show the truncated series expansion with $N=50$ for the inner part and $N=5, \ M=10$ in the outer part in green. The corresponding accuracy in the surface region is better than $10^{-3}$. 


\section{Applications}

\subsection{Using the Solution}
\label{AppUsing}

In the above discussion, we have solved the Gross-Pitaevskii-Poisson equations in Eq.~\ref{gpp-scaleinvariantgpp}, expressed in terms of the dimensionless variable $s$ and $v$. We have seen that the solution of these equations is fully parametrized by the dimensionless coupling parameter $\gamma$ and then obtained an analytic form for profile of boson stars. To use the solution to describe a boson star with mass $M$ made of boson with mass $m$ and self-coupling $\lambda$, we therefore need to obtain the corresponding value of the dimensionless coupling parameter $\gamma$. 

\begin{figure}[t]
\centering
\includegraphics[width=0.45\textwidth]{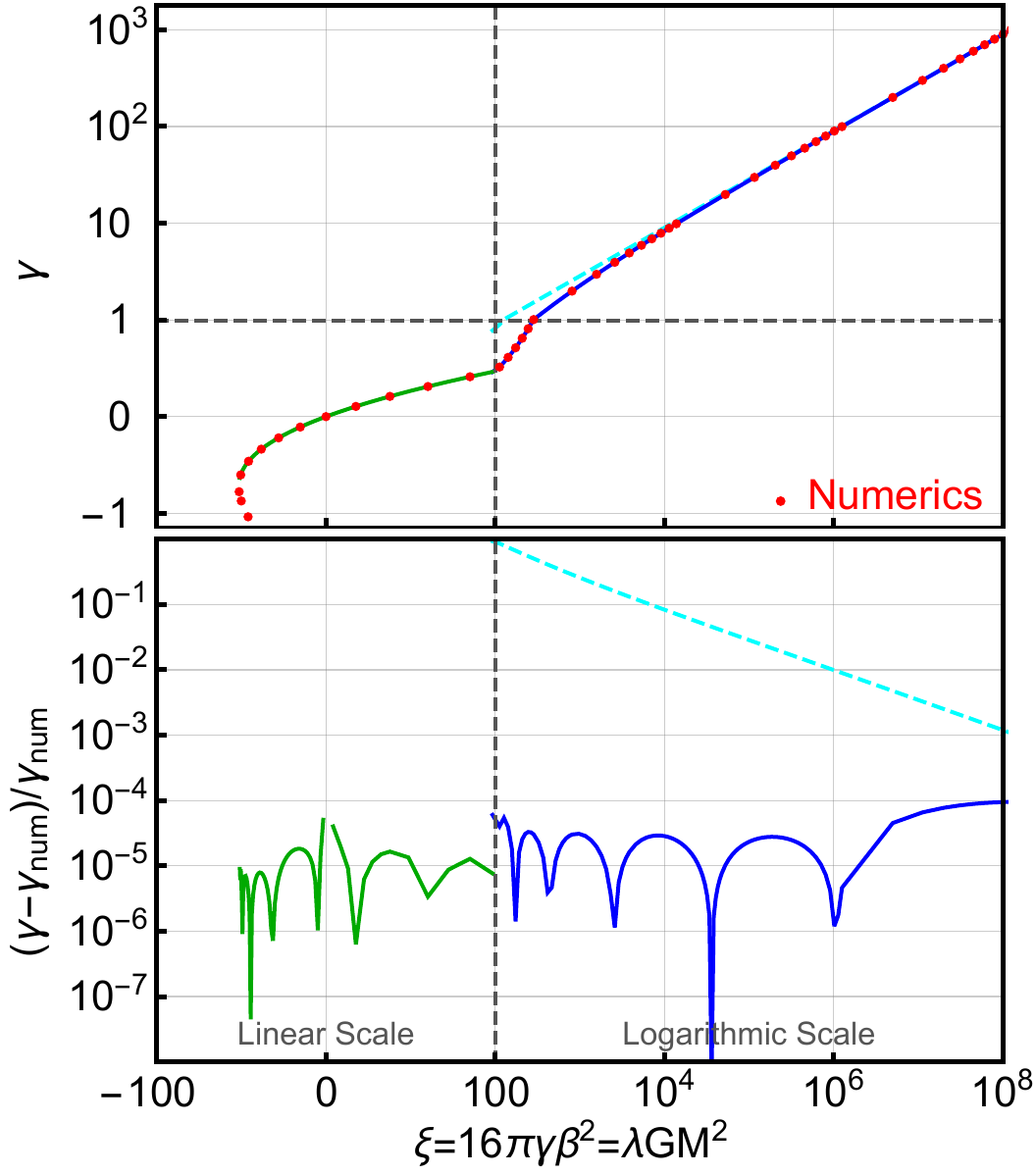}
\caption{The upper panels show the coupling parameter $\gamma$ as function of $\xi=\lambda G M^2$. The numerical solution is shown as red dots in the upper panels. The analytic fit to the numerical results is shown in green for the weak coupling regime (see Eq.~\ref{app:gb2-NI}), in blue for the strong coupling regime (see Eq.~\ref{app:gb2-TF}) and in cyan for the Thomas-Fermi limit. The lower panel shows the accuracy of the analytic solutions with respect to the numerical solution. 
}
\label{app:gammabeta2}
\end{figure}

Combining our scale choice $k=\frac{GMm}{2\beta}$ with Eq.~\ref{gpp-dimles-vars} and \ref{gpp-rescale}, we can see that
\be
\gamma = k^2 \Lambda = \frac{G M^2 \lambda }{16 \pi \beta^2 } .
\ee
Note that the value of the self-coupling parameter $\gamma$ is independent of the boson mass $m$ and only depends on the star's mass $M$ and the quartic self-coupling $\lambda$. We can obtain the coupling parameter $\gamma$ by solving 
\be
16 \pi \gamma \beta(\gamma)^2 = \lambda G M^2   \equiv  \xi
\label{app:eq-gammabeta2}
\ee
for $\gamma$ as a function of $\xi$ where we have introduced the short-hand notation $\xi=\lambda G M^2 $. Alternatively, we can follow our previous approach used to obtain the expansion coefficients and obtain an expression for $\gamma$ by finding a suitable fit to the numerical solution. This is shown in Fig.~\ref{app:gammabeta2}. The red dots correspond to the numerical solution using the same data set as Fig.~\ref{series:param-fit}. 

In the Thomas-Fermi limit, the expansion parameter $\beta$ is given by $\beta =\frac{\pi}{2} \sqrt{\gamma}$. We can solve Eq.~\ref{app:eq-gammabeta2} and obtain 
\be
\gamma^{TF} = \sqrt{\frac{\xi}{4 \pi^3 }}
\ee
as shown in  Fig.~\ref{app:gammabeta2} as dashed cyan curve. To obtain a more accurate result, we follow our approach used to obtain the expansion parameters and fit the numerical solution with a suitable series expansion. For large values of $\xi > 100$ we expand around the Thomas-Fermi limit and obtain
\begin{widetext}
\be
\hspace{-0.2cm}
\gamma(\xi) 
= 
\gamma^{TF}
\big[
1
 -0.035941 \ \xi^{-\frac{1}{4}}
 -9.569558 \ \xi^{-\frac{1}{2}}
+31.89268 \ \xi^{-\frac{3}{4}}
 -120.9668 \ \xi^{-1}
+316.0673 \ \xi^{-\frac{5}{4}}
 -308.2427 \ \xi^{-\frac{3}{2}}
\big]
\label{app:gb2-TF}
\ee
For small $\xi<100$ we find
\be
\gamma(\xi) 
= 
 -0.720960
&+1.157002  \left( \frac{\xi-\xi_{min}}{100} \right)^{\frac{1}{2}}
 -0.400828  \left( \frac{\xi-\xi_{min}}{100} \right) 
+0.420862  \left( \frac{\xi-\xi_{min}}{100} \right)^{\frac{3}{2}} \\
& -0.299337  \left( \frac{\xi-\xi_{min}}{100} \right)^{2} 
+0.125788  \left( \frac{\xi-\xi_{min}}{100} \right)^{\frac{5}{2}} 
 -0.023160  \left( \frac{\xi-\xi_{min}}{100} \right)^{3} 
\label{app:gb2-NI}
\ee
\end{widetext}
with $\xi_{min}=-51.523602$. 
Note that domain of $\xi$ is restricted to $ \xi>\xi_{min}$, and therefore, the coupling parameter $\gamma$ is bound from below $\gamma>\gamma_{min}=-0.722$. Physical boson stars therefore need to fulfill the condition $\lambda G M^2 > \xi_{min}$. This implies that the maximal mass of a boson star with attractive self-interaction, $\lambda<0$, is given by
\be
M_{max}^{\lambda<0} = M_{pl} \sqrt{\frac{\xi_{min}}{\lambda}}. 
\ee
There exists a second branch of the solution corresponding $\gamma<\gamma_{min}$. However, the corresponding boson star would have a higher total energy than the solution for $\gamma>\gamma_{min}$. Such configuration is unstable with respect to perturbations and therefore unphysical \cite{Chavanis:2011zi}. 

The accuracy of the expressions given in Eq.~\ref{app:gb2-NI} and Eq.~\ref{app:gb2-TF} with respect to the numerical solution is shown in the lower panel of Fig.~\ref{app:gammabeta2}. We can see that the analytic expressions reproduce the numerical results with accuracy better of $\mathcal{O}(10^{-4})$ for all physical values of $\xi$. \\

Knowing the value of $\gamma$ for a given boson star, we can obtain the profile for physical wave function $\psi$ and gravitational potential $\Phi$ by rescaling the dimensionless solution $s$ and $v$ according to Eq.~\ref{gpp-dimles-vars} and \ref{gpp-rescale}. We can write 
\be
\psi(r) &=  \sqrt{\frac{1}{8 \pi  }} \frac{G^{\frac{3}{2}} M^{\frac{3}{2}} m^3}{\beta^2} s\left( \frac{GMm^2 }{\beta}r\right)   \\
\Phi(r)&= -\frac{G^2 M^2 m^2}{2 \beta^2} \left[ 1+v\left( \frac{GMm^2 }{\beta}r \right)  \right]
\ee
Furthermore, we can simply read off the binding energy
\be
e= - 2m k^2 = -\frac{G^2 M^2 m^3}{2 \beta^2}
\label{app-binding}
\ee
and the central density
\be
\rho_0 = M \psi(0)^2 = \frac{1}{8\pi \beta^4}  G^3 M^4 m^6 s_0^2.
\ee

\subsection{Axion Stars} 

In the following, we illustrate the use of the obtained solution on one particularly well-motivated scenario: axion stars. The axion is a real pseudo-scalar field, which was initially introduced to solve the strong-CP problem. Furthermore it also provides a natural dark matter candidate if its mass in the range between $m=10^{-5}$~eV and $10^{-2}$~eV. The axion potential can heuristically be described by the instanton potential
\be
\hspace{-0.2cm}
V(a) 
= m_\pi^2 f_\pi^2 \left[1-\cos\left( \frac{a}{f}\right)\right] \approx \frac{1}{2} m^2 a^2 + \frac{\lambda}{12} a^4
\ee 
where $a$ denotes the axion field, $m_\pi=135$~MeV is the pion mass, $f_\pi=92$~MeV is the pion decay constant and $f$ is the axion decay constant. After expanding the potential and matching it to the form of Eq.~\ref{gpp-Lagrangian}, we find that the axion mass and decay constant are related by $m f = m_\pi f_\pi $.  The quartic coupling is given by $\lambda = - \frac{m^2}{ 2 f^2}$ where the negative sign indicates that the self-interaction is attractive. Note that there exist higher terms in the expansion. However, those terms will not contribute in the non-relativistic approximation and have been shown to be insignificant for dilute axions stars even in the relativistic limit \cite{Barranco:2010ib}. 

In the previous section, we have seen that a boson star with attractive self-interaction has an upper mass. Using the axion parameters this implies
\be
M<  10.14 \frac{m_\pi f_\pi }{ G^{\frac{1}{2}}m^2}  =2.74\cdot 10^{-12} M_\odot \left[\frac{10^{-5}~\text{eV}}{m}\right]^2.
\ee
We have seen in Sec.~\ref{gpp-validity} that the non-relativistic approximation is valid for axion stars with mass 
\be
M \ll \frac{1}{Gm}  = 5 \cdot 10^{-7} M_\odot \left[\frac{10^{-5}~\text{eV}}{m}\right].
\ee
For axions in the dark matter axion window of masses between $m=10^{-5}$~eV and $10^{-2}$~eV, the upper bound on the axion star's mass is well below this limit, and therefore, the axion star is well described by the non-relativistic approximation. 

\begin{figure}[t]
\centering
\includegraphics[width=0.45\textwidth]{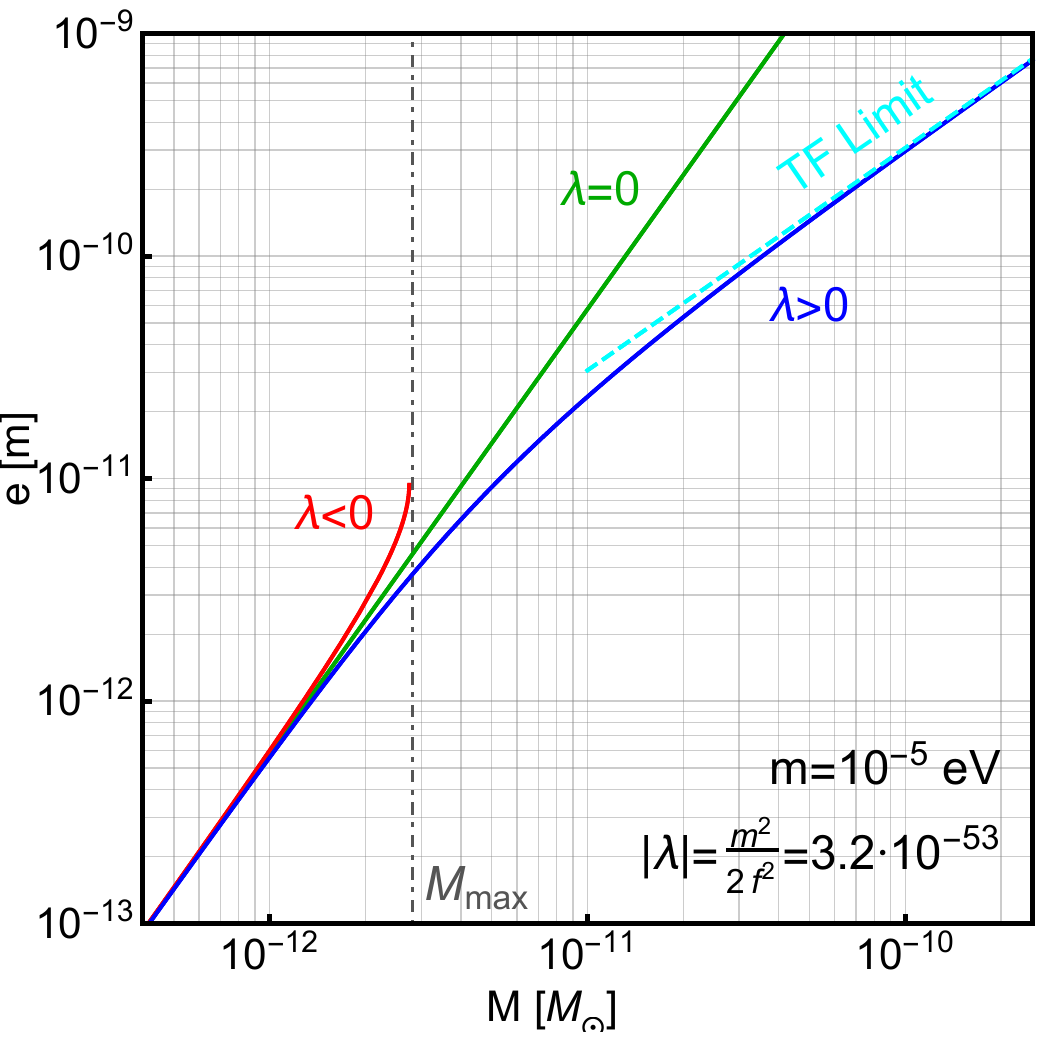}
\caption{Binding energy $e$ as a function of the axion star's mass $M$ for an axion mass $m=10^{-5}$~eV. We show the results for axion-star without self-interaction (green), with attractive self-interaction (red) and a \textit{wrong-sign} axion-star with a repulsive self-interaction (blue). The cyan line indicates the Thomas-Fermi limit. The vertical line indicates the upper bound on the star's mass $M_{max}$ in the case of an attractive self-interaction. }
\label{app:EvsM}
\end{figure}

In Fig.~\ref{app:EvsM} we show the binding energy $e$ given in Eq.~\ref{app-binding} as a function of the axion star's mass $M$ for a axion mass $m=10^{-5}$~eV. The green line shows the result for a non-interacting axion field, $\lambda=0$. In this case the binding energy is given by $e = - 0.1627 \cdot G^2 M^2 m^3$, which increases quadratically with the star's mass. The red lines indicates the axion field including the attractive self-interaction, $\lambda = - \frac{m^2}{ 2 f^2}= - 3.2 \cdot 10^{-53}$. For small axion star masses $M \lesssim 10^{-12}~M_\odot$, it approaches the non-interacting limit. In this case the classical pressure due to the self-interaction is negligible since the axion density is small even in the center of the star. For masses $M>10^{-12}~M_\odot$ the self-interaction term becomes important and the binding energy increases relative to the non-interacting case. The upper bound on the axion mass star mass, $M_{max}= 2.74\cdot 10^{-12} M_\odot$, is indicated by the gray dashed line. The authors of \cite{Enander:2017ogx} have shown that axions with mass $m=10^{-5}$~eV will form axion miniclusters with masses around $M \sim  10^{-12} M_\odot$ and therefore in the regime in which self-interactions are important. 

For comparison, we also show the results for a \textit{wrong-sign} axion star, in which the self-coupling has the same magnitude as the axion field but is repulsive, $\lambda = + \frac{m^2}{ 2 f^2} =+ 3.2 \cdot 10^{-53}$, as a blue line. At high masses, the binding energy approaches the Thomas-Fermi limit, shown as a dashed cyan line. In this case the binding energy is given by $e=-\frac{4}{\sqrt{\lambda \pi}} G^{\frac{3}{2}} M m^3$. 

\begin{figure}[t]
\centering
\includegraphics[width=0.45\textwidth]{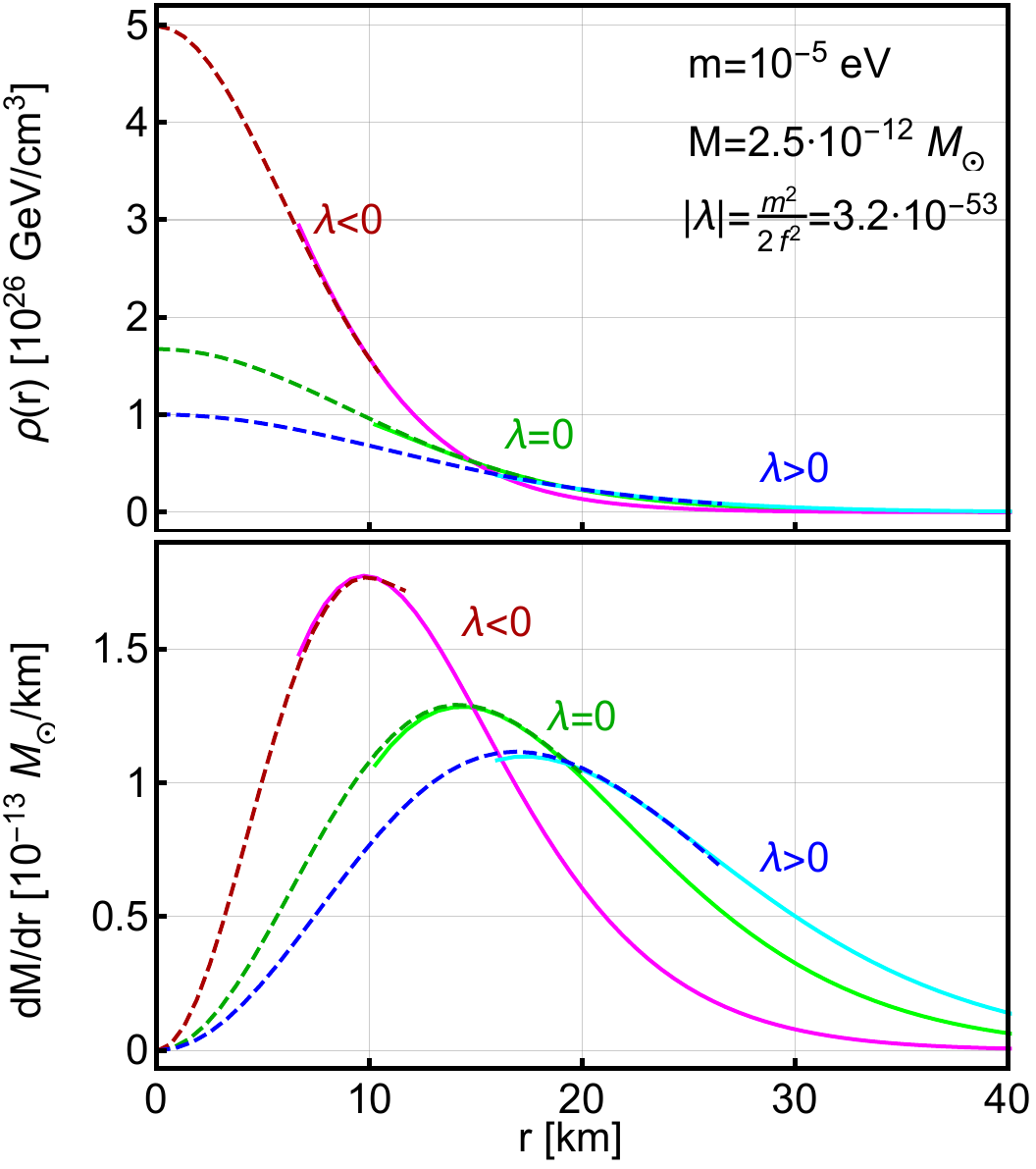}
\caption{The upper panel shows the energy density profile $\rho$ for an axion star with mass $M=2.5\cdot 10^{-12}~M_\odot$ and axion mass $m=10^{-5}$~eV. We show the results for an axion-star without self-interaction (green), with attractive self-interaction (red) and a \textit{wrong-sign} axion-star with a repulsive self-interaction (blue). We use the truncated series expansion of the wave function at small radius with $N=20$ (dashed) and with at large radius with $N=3, \ M=2$ (solid) as given in Eq.~\ref{series-truncated}. The lower panels shows the corresponding differential mass distributions $dM/dr$.}
\label{app:profile}
\end{figure}

In the upper panel of Fig.~\ref{app:profile} we show the density profile, $\rho(r) = M \psi^2(r)$. For illustration, we choose an axion mass  $m=10^{-5}$~eV and axion star mass $M=2.5\cdot10^{12}~M_\odot$ which is slightly below the maximal mass $M_{max}$. Following the previous discussion, we consider an axion-star without self-interaction (green), including the attractive self-interaction (red) and a \textit{wrong-sign} axion-star with a repulsive self-interaction (blue). To calculate the density profile, we use the truncated series expansion as given in Eq.~\ref{series-truncated} with $N=20$ at small radius, as indicated by the dashed lines, and $N=3, \ M=2$ at large radius as indicated by the solid lines. We can see that the two series expansions match well at intermediate values of the radius. Note that the attractive self-interaction of the axion field leads to a significant deformation of the density profile with respect to the a non-interaction axion star of same mass. This indicates the importance of including the axion self-interaction when considering axions stars close to their maximal mass $M \sim M_{max}$. 

The lower panel of Fig.~\ref{app:profile} shows the differential mass distribution $dM/dr = 4 \pi r^2 \rho$ for the three considered stars. We can see that the mass distribution peaks at intermediate radii and approximately half of the star's mass is described by both the small and large radius expansion of the wave function as given in Eq.~\ref{series-truncated}. This once again shows the importance of an accurate descriptions of the tails of the wave function profile in order to correctly describe the axion star.


\section{Conclusion}

Light scalar fields can form gravitationally bound compact objects, called boson stars. In the Newtonian limit, the profiles of boson stars are described by the Gross-Pitaevskii-Poisson equations. In a previous study~\cite{Kling:2017mif}, we presented a semi-analytic solution to these equations describing the profile of boson stars formed by a non-interacting scalar field. The solution is based on a series expansion which is parametrized by four expansion parameters that have been obtained from numerical simulation at high accuracy. 

In this study, we have generalized our semi-analytic approach to boson stars where the constituent particles have self-interactions. In this case, the expansion parameters are functions of the quartic self-coupling. Based on results from numerical simulations, we found a corresponding analytic expression for all expansion parameters. 

This allows to simply obtain profiles of boson stars in an analytic form for arbitrary self-couplings at high precision directly from the series expansion. In particular, no further time consuming and computational expansive numerical integration is needed.  

We have also applied our methods to axion stars, and shown how the the mass and density profiles can be obtained for both weak and strong interactions. The profiles are significantly modified from the case of non-interacting bosons. The methods developed in this paper allows for systematic studies of the properties of boson stars in an analytic way without further relying on numerical simulations. 

Finally, we note that there are several possible generalizations of these results. In particular, we can extend our results to rotating boson stars. It would also be interesting to see how the profiles are modified in the presence of other astrophysical objects like planets. We leave these and other questions  to future work. \\


\acknowledgments

We thank Joshua Eby and William Shepherd for useful discussions. This work is supported by NSF under Grant No. PHY-1620638. F.K. performed part of this work at the Aspen Center for Physics, which is supported by NSF Grant No. PHY-1607611.


\bibliographystyle{utphys}
\bibliography{references}

\end{document}